\documentclass[preprint,aps]{revtex4}
\usepackage[dvips]{graphicx}
\usepackage{color}
\newcommand{\blue}[1]{{\color{blue}}}
\begin{document}
\title{Levy-Lieb principle: The bridge between the electron density of Density Functional Theory and the wavefunction of Quantum Monte Carlo\footnote{Chemical Physics Letters, in press (2015)}}
\author{Luigi Delle Site}
 \email{luigi.dellesite@fu-berlin.de}
\affiliation{Institute for Mathematics, Freie Universit\"{a}t Berlin\\
Arnimallee 6, D-14195 Berlin, Germany.}

\begin{abstract}
The constrained-search principle introduced by Levy and Lieb, is proposed as a practical, though conceptually rigorous, link between Density Functional Theory (DFT) and Quantum Monte Carlo (QMC). The resulting numerical protocol realizes in practice the {implicit} key statement of DFT:\\
``{\it Given the three dimensional electron density of the ground state of a system of $N$ electrons with external potential $v({\bf r})$ it is possible to find the corresponding $3N$-dimensional wavefunction of ground state.}''\\
From a numerical point of view, the proposed protocol can be employed to speed up the QMC procedure by employing DFT densities as a {\bf pre-selection} criterion for the sampling of wavefunctions.
\end{abstract}
\maketitle
\section{Introduction}
The Hohenberg-Kohn (HK) formulation of DFT and the subsequent development of the
Kohn-Sham (KS) scheme made DFT the most popular tool for electronic structure calculations
(see, e.g. \cite{ref1}). The {profound} meaning of the HK theorems for many-electron systems {was further strengthened} by the Levy-Lieb formulation of the problem \cite{ref2,ref3}. In the next section the relevant aspects of the Levy-Lieb formulation will be reported. {Such a formulation removes limitations of the HK theorem, such as the degeneracy of the ground state, and  sets the exact correspondence between the 3-dimensional electron density and the 3N-dimensional wavefunction of the
ground state of a system of $N$ electrons}. In practical calculations, this formulation is never explicitly used and {it is usually considered only a conceptual proof of validity of DFT}. In this paper I present the Levy-Lieb formulation in a different perspective; I propose such a formulation as the central algorithm {which} consistently merge two complementary but distinguished, approaches, namely KS DFT and QMC. The merging occurs in such a way that the numerical efficency of DFT contributes to enhance the numerical efficency of QMC and that the poor accuracy of DFT is enhanced by the accuracy of QMC.
A similar idea, restricted to Orbital-Free DFT and to Ground State  Path Integral QMC was presented in a previous paper \cite{ijqc}; here the idea is extended to all QMC methods based on the explicit calculation of the ground state wavefunction (above all the most recent characterized high accuracy of the wavefunction e.g. \cite{mor,ali}), and to the KS DFT scheme. {Another (suggestive, but yet not fully proved) implication of the procedure proposed is that the Levy-Lieb principle can be interpreted as a particular case of a decoding key of a set of 3-dimensional data into a set of $3N$-dimensional data within the more general framework of Information Theory \cite{ijqcit}; this interpretation may have important implications for problems of ``inverse chemistry'' (see e.g. Ref.\cite{reiher}). However in this paper I wish only to mention the possible ``decoding'' interpretation and a detailed discussion can be found in Ref.\cite{ijqcit}.}
\section{Levy-Lieb formulation and the exchange-correlation functional}
 The Levy-Lieb constrained-search principle is formulated as:
\begin{equation}
E_{gs}=Min_{\rho}\left[Min_{\psi\to \rho}\left<\psi\left|\widehat{T}+\widehat{V}_{ee}\right|\right>+\large\int v({\bf r})\rho({\bf r})d{\bf r}\right].
\label{eq1}
\end{equation}
with $E_{gs}$ the ground state energy, $\psi({\bf r}_{1}......{\bf r}_{N})$ an antisymmetric wavefunction of an $N$-electron system, $\widehat{T}$ the kinetic operator, $\widehat{V}_{ee}$ the electron-electron Coulomb potential operator and $v({\bf r})$ the external potential (usually electron-nucleus Coulomb potential). 
The inner minimization is restricted to all antisymmetric wavefunctions $\psi$ leading to $\rho({\bf r})$, i.e. $\rho({\bf r})=N\large\int_{\Omega_{N-1}}\psi^{*}({\bf r},{\bf r}_{2},.....{\bf r}_{N})\psi({\bf r},{\bf r}_{2},.....{\bf r}_{N})d{\bf r}_{2}....d{\bf r}_{N}$ (where $\Omega_{N-1}$ is the $N-1$ spatial domain), while the outer minimization is then independent of $\psi$ and searches over all the $\rho$'s which integrate to $N$. Within this framework, the universal functional of the Hohenberg-Kohn formulation of DFT  is written as:
\begin{equation}
F[\rho]=Min_{\psi\to \rho}\left<\psi\left|\widehat{T}+\widehat{V}_{ee}\right|\psi\right>.
\label{eq2}
\end{equation}
The universal functional $F[\rho]$ is unknown, however the accuracy of its approximations is the key to the accuracy of DFT calculations. More specifically, within the KS formulation of DFT, that is the most widely used approach of DFT for electronic structure calculations, it is not the entire $F[\rho]$ that needs to be approximated, but only the so called exchange and correlation part of it, $E_{xc}[\rho]$. Consistently with the definition of $F[\rho]$, $E_{xc}[\rho]$ can be defined as follows:
\begin{equation}
E_{xc}[\rho]=F[\rho]-T_{nint}[\rho]-E_{H}[\rho]
\label{eq3}
\end{equation} 
where $T_{nint}[\rho]$ is the kinetic functional for non-interacting particles (whose explicit expression is also unknown, see e.g. \cite{trick})and $E_{H}[\rho]=\int\frac{\rho({\bf r})\rho({\bf r}^{'})}{|{\bf r}-{\bf r}^{'}|}d{\bf r}d{\bf r}^{'}$ is the Hartree energy.

\section{Speed up the initial stage of QMC by sampling $\psi$ at given $\rho$}
\begin{figure}
\center
\includegraphics[width=0.95\textwidth]{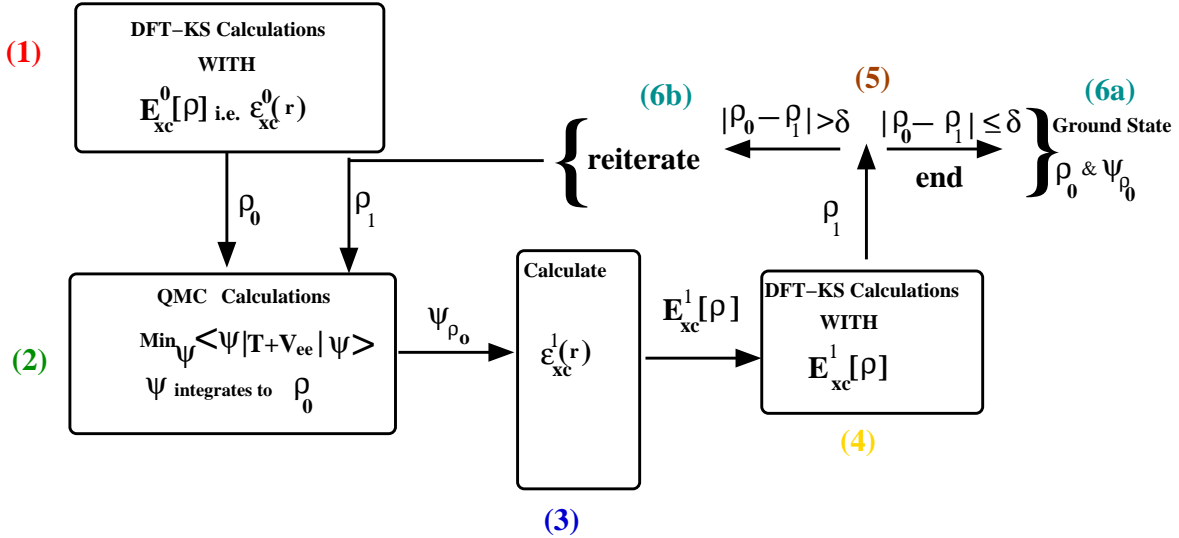}
\caption{The schematic representation of the iterative procedure.\label{fig1}}
\end{figure}

Let us suppose we employ a given expression of $E_{xc}[\rho]$ (let us refer to it as: $E^{0}_{xc}[\rho]=\int\rho({\bf r})\epsilon^{0}_{xc}({\bf r})d{\bf r}$) in a DFT-KS calculation with external potential $v({\bf r})$. The KS calculation done with $E^{0}_{xc}[\rho]$ will deliver a density of ground state $\rho_{0}({\bf r})$. In parallel, let us consider the problem: $Min_{\psi}\left<\psi\left|\widehat{T}+\widehat{V}_{ee}\right|\psi\right>$ within a QMC approach. The search for $\psi^{min}$, in this case, would correspond to find the ground state wavefunction of a gas of $N$ electrons, that is we sample the space of antisymmetric $\psi$'s, and a wavefunction is preferred to another if its energy is lower in absence of any other external constraints. However, if, within the QMC procedure, we add the constraint that $\psi$ must integrate to a given $\rho({\bf r})$, (e.g. to $\rho_{0}({\bf r})$), then we numerically realize the Levy-Lieb principle, that is we practically realize the operation of $Min_{\psi\to \rho}$ of Eq.\ref{eq1}.\\
The relevant, and non trivial, consequence of the argument above is that, if  $E^{0}_{xc}[\rho]$ was exact, then $\rho_{0}({\bf r})$ would be exact and the $\psi$ delivered by QMC as a solution of $Min_{\psi}\left<\psi\left|\widehat{T}+\widehat{V}_{ee}\right|\psi\right>$ with the constraints that $\psi$ integrates to $\rho_{0}({\bf r})$, would correspond to the exact (within the accuracy of QMC) wavefunction of the ground state for a system of $N$ electrons with external potential $v({\bf r})$. Based on the considerations above, the proposed procedure to speed up QMC,  consists of the following steps:
\begin{itemize}
\item Employ $E^{0}_{xc}[\rho]$ for a KS calculation and obtain $\rho_{0}({\bf r})$.
\item Perform a QMC calculation where $\psi$ is sampled in such a way that it integrates to $\rho_{0}({\bf r})$ and minimizes $\left<\psi\left|\widehat{T}+\widehat{V}_{ee}\right|\psi\right>$. Let us call the corresponding wavefunction $\psi_{\rho_{0}}$.
\item If the DFT functional employed is accurate enough, then $\psi_{\rho_{0}}$ would represent a very good {\bf initial guess} for solving the full problem in QMC.
\end{itemize}
The essence of the scheme is that one, computationally trivial, DFT calculation can {\bf always} be used to simplify a complex QMC calculation,
at least in its initial stage. 
{The practical question concerns the construction of} an efficient numerical algorithm for sampling $\psi$ {\bf at given} $\rho({\bf r})$ (suggestions for a specific case are given in Ref.\cite{book-ch}). Here for efficiency is meant that the computational cost for the convergence of the sampling at given $\rho({\bf r})$ must be negligible compared to the computational cost of convergence of the standard QMC procedure for searching $\psi$. {The suggested procedure is well defined in each of its step and its success/failure depends only on the numerical efficiency  in sampling $\psi$ {\bf at given} $\rho({\bf r})$. However, one may aim to extend the procedure beyond its utility for the ``initial stage'' of a QMC calculation and ask whether it may be possible to design a combined KS DFT-QMC scheme to iteratively solve the ground state problem; in the next section I propose an iterative scheme.}
{\section{Iterative DFT-QMC procedure: self-consistent derivation of the ground state}}
The key issue for the efficiency of the scheme suggested in the previous section is the accuracy of  $E^{0}_{xc}[\rho]$, the more accurate is $E^{0}_{xc}[\rho]$ the closer $\psi_{\rho_{0}}$  from QMC is to the real ground state wavefunction.
{In general $E^{0}_{xc}[\rho]$ may not be enough accurate and in this section I suggest a procedure that updates $E_{xc}[\rho]$ and $\psi$ in iterative manner. Interestingly, upon convergence, the procedure goes beyond the ``initial stage'' of a QMC calculation and leads, in principle, to the ground state wavefunction.}
For this purpose I define:
\begin{equation}
F[\rho]=\int \rho({\bf r})f({\bf r})d{\bf r}.
\label{frho}
\end{equation}
{Next I need to introduce a ``universal'' kinetic and Hartree energy reference for non interacting electrons. When this reference energy is subtracted to $F[\rho]$ one obtains the exchange and correlation energy as in Eq.\ref{eq3}.} 
\begin{equation}
E_{H}[\rho]=\int \rho({\bf r})e_{H}({\bf r})d{\bf r};e_{H}({\bf r})=\int\frac{\rho({\bf r}^{'})}{|{\bf r}-{\bf r}^{'}|}d{\bf r}^{'} 
\label{ehrho}
\end{equation}
\begin{equation}
T_{nint}[\rho]=\int \rho({\bf r})t_{nint}({\bf r})d{\bf r}
\label{Tninit}
\end{equation}
where  
\begin{equation}
t_{nint}({\bf r})=t_{HF}({\bf r}) 
\label{thf}
\end{equation} 
$HF$ stays for Hartree-Fock, $t_{HF}({\bf r})=\frac{K_{HF}({\bf r})}{\rho({\bf r})}$, and $K_{HF}({\bf r})$ is such that the total kinetic energy writes, $T_{HF}=\int K_{HF}({\bf r}) d{\bf r}$. 
This means that the kinetic energy density for non interacting electrons is determined by the Hartree-Fock calculation of the $N$-electron system considered. 
For consistency with the definition of $t_{nint}({\bf r})$ we define:
\begin{equation}
E_{H}[\rho]=\int \rho({\bf r})e^{HF}_{H}({\bf r})d{\bf r}
\label{ehrho1}
\end{equation}
that is the Hartree energy density is calculated by the Hartree-Fock method for the specific $N$-electron system considered. In this way we consider $t_{nint}({\bf r})$ and $e^{HF}_{H}({\bf r})$ as reference quantities in the limit in which electron correlations, due to the explicit electron-electron interaction, are absent. {The reason for calculating $e_{H}({\bf r})$ and $t_{nint}({\bf r})$ with HF is due to the need of having a general reference energy for non interacting particles. Such a quantity shall be universal and independent of a specific $\rho({\bf r})$ generated by a specific choice of a $E_{xc}[\rho]$, otherwise the iterative procedure shown below would be biased by the specific choice of the initial $E_{xc}[\rho]$.}
It must be noticed that { the exchange term of the HF calculation is not considered and used since this latter is automatically calculated in $f({\bf r})$ within the QMC approach.}
{The HF calculation is not needed for the first step of the procedure (``initial stage'' of the QMC calculation) outlined in the previous section, but, as it is reported next, it is required to update/improve $E^{0}_{xc}[\rho]$ at the successive steps. In any case the HF calculation is needed only once and is numerically not expensive.}
The iterative procedure consists of the following steps (pictorially illustrated in Fig.\ref{fig1}):
\begin{itemize}
\item Employ a chosen $E^{0}_{xc}[\rho]$ for a KS calculation and obtain $\rho_{0}({\bf r})$.
\item Perform a QMC calculation where $\psi$ is sampled via minimization of $\left<\psi\left|\widehat{T}+\widehat{V}_{ee}\right|\psi\right>$ with the additional request that $\psi$ integrates to  $\rho_{0}({\bf r})$, let us indicate the resulting wavefunction as $\psi_{\rho_{0}}$.
\item From $\psi_{\rho_{0}}$ we calculate $f^{0}({\bf r})$:
\begin{equation}
f^{0}({\bf r})=\frac{1}{\rho_{0}({\bf r})}\int\left<\psi_{\rho_{0}}\left|\widehat{T}+\widehat{V}_{ee}\right|\psi_{\rho_{0}}\right>d{\bf r}_{2}....d{\bf r}_{N}
\label{f0}
\end{equation}
\item At this point we can numerically update the exchange and correlation energy density:
\begin{equation}
\epsilon^{1}_{xc}({\bf r})=f^{0}({\bf r})-e_{H}({\bf r})-t_{nint}({\bf r})
\label{epsexc}
\end{equation}
{However rather than $\epsilon^{1}_{xc}({\bf r})$ what is needed is the exchange and correlation potential: $v_{cx}=\frac{\delta E_{xc}[\rho({\bf r})]}{\delta\rho({\bf r})}$. In order to have $v_{xc}$, one possibility is that of fitting numerically $\epsilon^{1}_{xc}({\bf r})$ in terms of $\rho^{0}({\bf r})$. The numerical fit delivers an analytical expression for $\epsilon^{1}_{xc}(\rho^{0}({\bf r}))$. The analytic expression is taken as an approximation of $\epsilon^{1}_{xc}(\rho({\bf r}))$ and can be employed in the KS procedure to derive (analytically) $v_{xc}$. The fitting procedure could be performed, for example, with approaches similar to the ``kernel ridge regression'' (KRR) a method for non-linear regression that prevents over-fitting \cite{krr} and that is successfully employed in machine learning approaches (see e.g.\cite{kieron1,kieron2,anatole1,anatole2})}.
\item  We can now employ $\epsilon^{1}_{xc}(\rho({\bf r}))$ (or better the corresponding $v_{xc}$) for a KS calculation which will deliver an updated electron density $\rho^{1}({\bf r})$.\\  
Next, starting from $\rho^{1}({\bf r})$ the scheme above can be repeated and continued until certain criterion of convergence on $\rho({\bf r})$, chosen a priori, is satisfied.
\item {Once the procedure converges,} the QMC wavefunction obtained at the final $i-th$ iteration, $\psi_{\rho_{i}}$, corresponds to the ``true'' ground state wavefunction of the system.
\end{itemize}

{Finally it must be reported that here the convergence of the iterative procedure, has not been proved mathematically. However the idea remains valid as long as a check that the total energy decreases at each iterative step is performed. As long as the total energy decreases, the DFT-QMC procedure is useful and at any successive step brings us closer to the exact ground state. If the energy increases at the $i$-th iteration step, then, the wavefunction, $\psi_{i-1}$, of the $i-1$-th iteration step shall be used as initial guess for a standard full QMC calculation.} 
\section{Conclusion}
We have proposed a scheme to link in a consistent way KS-DFT and QMC via the Levy-Lieb principle. The electron density of KS-DFT, that is a quantity solution of a $3$-dimensional problem can be used to significantly restrict the $3N$-dimensional space of wavefunctions sampled by the QMC procedure.{ The procedure can be divided in two parts: first, given any exchange and correlation functional, the corresponding electron density, solution of the KS equation, can be used as a pre-selection criterion for sampling a wavefunction that can be employed as initial guess in a QMC calculation. One could stop here and by employing such initial guess proceed with the standard full QMC procedure. The second part of the scheme proposed is a continuation of the first part, that is a procedure to iteratively solve the ground state problem by combining at each step KS calculations and subsequent QMC sampling. This second part requires an additional HF calculation, in order to define a general energy of reference for non interacting electrons, and a numerical fitting procedure}. 
It must be underlined that the whole procedure is based on the assumption that there exists an efficient numerical procedure of sampling $\psi$ at a given electron density $\rho$; despite some proposals were put forward in previous work, up to now there is no numerical implementation of such a scheme. {Moreover the application of the KRR method for the fitting procedure may require a considerable amount of computational resources, thus its application may also need to be optimized so that the total cost of the proposed scheme is convenient compared to that of a full QMC calculation}. Thus this work must be considered as a protocol that may be worth to explore further and eventually implement in a computational code.
\section{Acknowledgments} 
{I would like to thank Luca Ghiringhelli for a critical reading of the manuscript, for the clarifying lecture about the KRR method and for several valuable suggestions. I would also like to thank Anatole von Lilienfeld for providing references for the KRR method and its applications. This work was supported by the Deutsche Forschungsgemeinschaft (DFG), via the Heisenberg
Grant, No. DE 1140/5-2.}   

\end{document}